\documentclass[fleqn,twoside]{article}%
\usepackage{amsmath}
\usepackage{amsfonts}
\usepackage{espcrc2}
\usepackage[figuresright]{rotating}
\usepackage{amssymb}
\usepackage{graphicx}%
\begin{document}

\title{MiniBooNE and Sterile Neutrinos}%
\author{M. H. Shaevitz\address
{Columbia University,  Department of Physics, New York, NY 10027}%
\thanks{shaevitz@nevis.columbia.edu}
for the BooNE Collaboration[1]\thanks{This work supported by the NSF and DOE.}%
}%

\begin{abstract}%
Sterile neutrinos may be an important extension to the standard model, and
could both hold the key to understanding neutrino mass and mixing as well as
play an important role in leptogenesis. In many models, the sterile neutrinos
could be light and accessible to current and near term experiments. The
MiniBooNE experiment is set up to explore these possibilities in the $\Delta
m^{2}$ region from 0.3 to a few eV$^{2}$ where the LSND experiment has
reported a $\overline{\nu}_{e}$ appearance signal. This report will outline
some of these extensions, give the status and prospects for the MiniBooNE
experiment, and explore future investigations if MiniBooNE sees an oscillation
signal.
\end{abstract}%

\maketitle

\section{Extensions to the Neutrino Standard Model}

The standard model has three flavors of massless Dirac neutrinos $\nu_{e}%
,\nu_{\mu},\nu_{\tau}$. The results of neutrino oscillation measurements over
the last decade have required that this simple picture be modified to allow
mixing between the flavors and to add mass for the neutrinos. These
modifications have opened up many questions that are currently being pursued
experimentally. The questions include: What are the number of neutrinos? Are
the masses Majorana or Dirac? Does the neutrino mass hierarchy follow that of
the charged leptons? Why are the mixings so different from the quark sector?

\subsection{Extensions with Sterile Neutrinos}

The current experimental measurements in the solar ($\Delta m_{Solar}%
^{2}\approx7\times10^{-5}$ eV$^{2}$) and atmospheric ($\Delta m_{Atmosph.}%
^{2}\approx2.5\times10^{-3}$ eV$^{2}$) regions combined with the LSND result
($\Delta m_{LSND}^{2}=0.3-3$ eV$^{2}$) give three disparate mass squared
differences. The three $\Delta m^{2}$ cannot be explained with only three
types of neutrinos since
\[
\Delta m_{LSND}^{2}\neq\Delta m_{Solar}^{2}+\Delta m_{Atmosph.}^{2}%
\]
Only-three models, where the atmospheric oscillations correspond to a mixture
of oscillations at the $\Delta m_{Solar}^{2}$ and $\Delta m_{LSND}^{2}$
scales, are strongly disfavored by the current data since these models would
demand a strong $\nu_{e}$ appearance signal that is not seen in the Super-K
atmospheric data.

The solution to this inconsistency opens many possibilities. First, one of the
experimental measurements could be wrong. The recent results from the SNO and
Kamland experiments have substantiated the oscillation interpretation of the
solar neutrino deficit, and the K2K experiment, although with large
uncertainties, has seen a disappearance signal in the Super-K atmospheric
region. MiniBooNE is in the process of exploring oscillations in the LSND
region and will make a definitive check of the LSND anomaly.

Another possibility is that one of the experiments is not seeing neutrino
oscillations but the results of another type of process such as neutrino
decay; or, for example, for LSND, neutrino production from a lepton flavor
violating decay. It also may be possible that this inconsistency has its
origin in CPT violation which could imply that oscillations may be different
for neutrinos and antineutrinos. The recent confirmation by Kamland of
oscillations in the solar region using reactor antineutrinos limits any CPT
interpretation for the solar region. In the future, the MINOS experiment will
be able to check CPT invariance for the oscillations in the atmospheric region
by performing separate measurements with neutrinos and antineutrinos.

An elegant solution that could explain the three distinct mass differences is
the addition of one or more sterile neutrinos. Sterile neutrinos are natural
extensions to the standard model, and are demanded in many grand-unified
theories, extra-dimension theories, and various types of lepton symmetry
models. A confirmation that sterile neutrinos exist would have a major impact
on our understanding of the neutrino sector and have far reaching effects on
building models from unification to cosmology. Also, additional sterile
neutrinos could be a significant source of measurable CP violating effects and
will make the neutrino oscillation phenomenology much more diverse.

\begin{figure}
[ptb]
\begin{center}
\includegraphics[
height=1.5835in,
width=3.039in
]%
{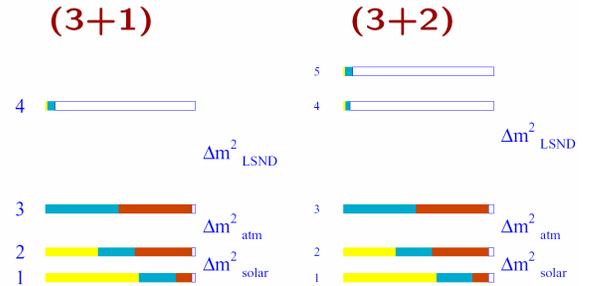}%
\caption{Example flavor compositions of the neutrino mass eigenstates for a
3+1 and 3+2 model with extra sterile neutrinos. The composition of each mass
state is given by the colored bars with electron-type (yellow), muon-type
(blue), tau-type (red), and sterile-type (white).}%
\label{3plusN}%
\end{center}
\end{figure}

\begin{figure}
[ptb]
\begin{center}
\includegraphics[
height=1.0568in,
width=3.039in
]%
{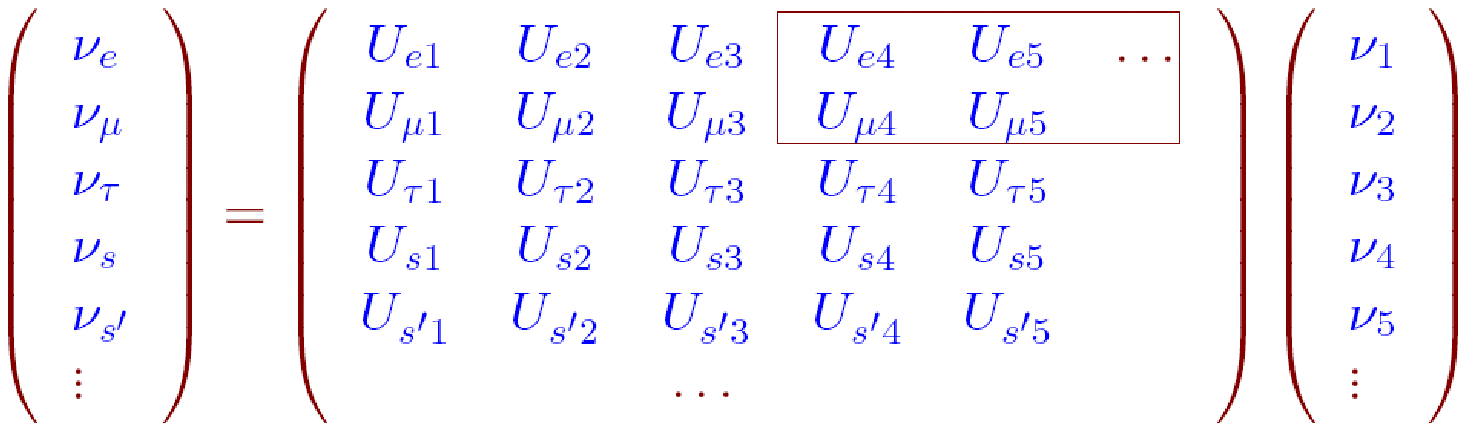}%
\caption{The neutrino mixing matrix for a 3+n model with extra sterile
neutrinos, $\nu_{s}$ and $\nu_{s^{\prime}}.$}%
\label{3plusNMatrix}%
\end{center}
\end{figure}

The phenomenology for adding extra sterile neutrinos involves broadening the
mixing matrix and adding new extra mass values. The sterile neutrinos are weak
isospin singlets and thus do not couple to either the standard W or Z bosons.
Sterile neutrinos would therefore evade the LEP bound on the number of light
active neutrinos, and the detection of sterile neutrino effects could only
take place through mixing with the standard active neutrinos. In the 3+1 and
3+2 schemes, additional sterile states are added with masses corresponding to
the $\Delta m_{LSND}^{2}$ region. The high mass states are mainly composed of
sterile components but can have small admixtures of active states as shown in
Figure~\ref{3plusN}. The extended neutrino mixing matrix is shown in
Figure~\ref{3plusNMatrix}, where the upper left $3\times3$ components
correspond to the usual active neutrino mixing matrix and the $U_{e4(5)}$ and
$U_{\mu4(5)}$ components would contribute to an LSND $\nu_{e}$ appearance
signal through oscillation through the sterile neutrinos.%

A sterile neutrino has also been proposed to explain some small solar neutrino
discrepancies such as the argon neutrino rate being two sigma smaller than
expected and the lack of an upturn in the Be neutrino rate at low
energies\cite{SmirnovSterile}. Sterile neutrinos may also be important in
contributing to the rapid-neutron-capture process (r-process) in heavy element
production by supernovae\cite{FullerRprocess}. The observed abundance of heavy
elements is much larger than standard model predictions where the heavy
element production is limited by the small size of the the neutron density.
The neutron density can be increased in a model where matter effects cause
more $\nu_{e}$'s to oscillate to sterile neutrinos than $\overline{\nu}_{e}%
$'s. The excess of the $\overline{\nu}_{e}$'s then produces a substantial
neutron excess through the inverse beta decay process.

Big-bang nucleosynthesis and the anisotropy measurements of the cosmic
microwave background give constraints on the number of neutrinos including
sterile neutrinos. Standard cosmological models using current measurements
constrain the total number of neutrinos to be $2.6\pm0.4$ which can be
somewhat relaxed to $4.0\pm2.5$ if the systematic uncertainty on the $^{4}$He
abundance is larger. These cosmological constraints can be evaded if the
neutrinos have a lepton asymmetry or if equilibrium assumptions are not
valid\cite{NuNumCosmo}. In fact, Hannestad has claimed that \textquotedblleft
the LSND result is not yet ruled out by cosmological
observations\textquotedblright\ and has shown how the cosmological mass bounds
would be increased from $\Sigma m_{\nu_{i}}=1.0$ to $1.4$ $(2.5)$ eV if there
are four (five) neutrinos instead of the usually assumed three\cite{Hannestad}.

\subsection{3+1 and 3+2 Global Fits to Neutrino Oscillation Results}

There have been many searches for neutrino oscillations in the high mass
$\Delta m^{2}$ region associated with the LSND result including the KARMEN,
NOMAD, Bugey, CHOOZ, CCFR84, and CDHS experiments. These experiments have set
both appearance and disappearance limits on oscillations at various levels.
Global fits\cite{Sorel3plus2} to these null short-baseline (NSBL) experiments
plus the LSND signal have been performed to determine the allowed regions in
oscillation parameter space and answer the question if there are regions of
compatibility. In these fits, LSND is the only positive signal (at the 3.8
$\sigma$ level), but the CDHS experiment also shows a two sigma deficit in the
near detector.

\begin{figure}
[ptb]
\begin{center}
\includegraphics[
height=3.5483in,
width=3.039in
]%
{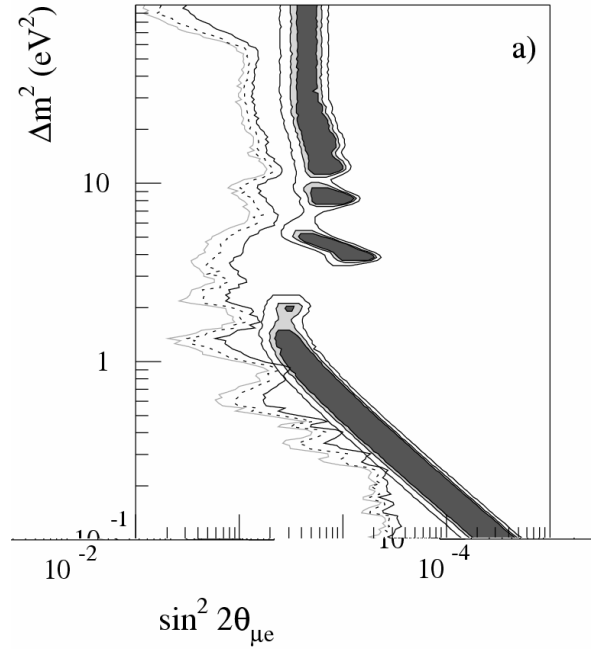}%
\caption{Compatibility between NSBL and LSND datasets in (3+1) models. The
figure shows the 90\% (grey solid line), 95\% (black dotted line), and 99\%
(black solid line) CL exclusion curves ($\sin^{2}2\theta_{\mu e},\Delta m^{2}%
$) space for (3+1) models, considering the NSBL experiments. Also shown are
the 90\%, 95\%, and 99\% CL allowed regions for the LSND data.}%
\label{3Plus1overlap}%
\end{center}
\end{figure}

\begin{figure}
[ptb]
\begin{center}
\includegraphics[
height=3.2707in,
width=3.039in
]%
{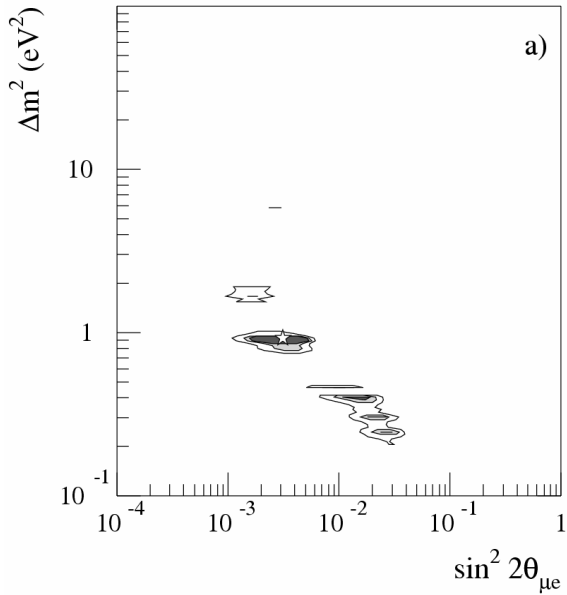}%
\caption{Allowed regions in parameter space from a combined analysis of NSBL
and LSND data, in (3+1) models, assuming statistical compatibility of the NSBL
and LSND datasets. The figure shows the 90\%, 95\%, and 99\% CL allowed
regions in ($\sin^{2}2\theta_{\mu e},\Delta m^{2}$) space, together with the
best-fit point, indicated by the star.}%
\label{3plus1}%
\end{center}
\end{figure}

From fits to 3+1 models, a compatibility (calculated from $1-\delta
_{\hbox{\tiny
NSBL}}(\delta_{\hbox{\tiny LSND}}+(1-\delta_{\hbox{\tiny LSND}})/2)$) of 3.6\%
between LSND and the NSBL experiments is obtained with the overlap regions
shown in Figure~\ref{3Plus1overlap}. ($\delta_{\hbox{\tiny
NSBL}}$ and $\delta_{\hbox{\tiny LSND}}$ are the confidence levels of the fit
for the NSBL and LSND data respectively.) This value of compatibility does not
support any conclusive statement, although it represents poor agreement
between the LSND and NSBL data sets. Assuming compatibility, the allowed
regions in parameter space are shown in Figure~\ref{3plus1} with best fit
values for the parameters of $\Delta m_{41}^{2}=0.92$ eV$^{2}$ with
$U_{e4}=0.136$ and $U_{\mu4}=0.205$. The global $\chi^{2}$ minimum is
$\chi_{\hbox{\tiny SBL}}^{2}$=144.9 (148 d.o.f.) with the individual NSBL and
LSND contributions being $\chi_{\hbox{\tiny SBL}}^{2}$ =137.3 and
$\chi_{\hbox{\tiny LSND}}^{2}$=7.6, respectively.%

\begin{figure}
[ptb]
\begin{center}
\includegraphics[
height=2.5598in,
width=3.039in
]%
{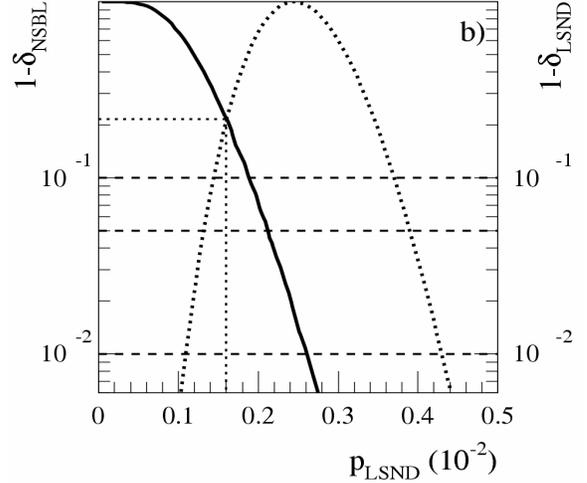}%
\caption{Individual confidence levels $\delta_{\hbox{\tiny NSBL}}$ and
$\delta_{\hbox{\tiny LSND}}$, as a function of the LSND oscillation
probability $p_{\hbox{\tiny LSND}}$, for the NSBL and LSND datasets. The
curves are for (3+2) models with the neutrino mass splittings, $\Delta
m_{41}^{2}$ and $\Delta m_{51}^{2}$, fixed to the best-fit values $\Delta
m_{41}^{2}=0.92\ \hbox{eV}^{2}$, $\Delta m_{51}^{2}=22\ \hbox{eV}^{2}$ from
the combined NSBL+LSND analysis, and variable mixing matrix elements $U_{e4}$,
$U_{\mu4}$, $U_{e5}$, $U_{\mu5}$. The left curve refers to the NSBL dataset,
the right one to the LSND dataset. }%
\label{3plus2pLSND}%
\end{center}
\end{figure}

The (3+2) model fits give much better compatibility than the (3+1) fit, beyond
what would be the statistical expectation for adding an additional three
parameters associated with the second sterile neutrino, $\Delta m_{51}^{2}$,
$U_{e5}$ and $U_{\mu5}$. Since two parameter representations in terms of
$\sin^{2}2\theta_{\mu e}$ and $\Delta m^{2}$ are not possible in a (3+2)
model, the results of the 3+2 fits are displayed in terms of the parameter,
$p_{LSND}$, the probability of giving $\nu_{\mu}\rightarrow\nu_{e}$
transitions for the LSND L/E distribution. From Figure~\ref{3plus2pLSND} which
shows a maximum compatibility at $\delta_{\hbox{\tiny NSBL}}=\delta
_{\hbox{\tiny LSND}}=0.785$, one obtains a compatibility value of 30\% as
compared to the 3.6\% for the (3+1) models.

\begin{figure}
[ptb]
\begin{center}
\includegraphics[
height=3.039in,
width=3.039in
]%
{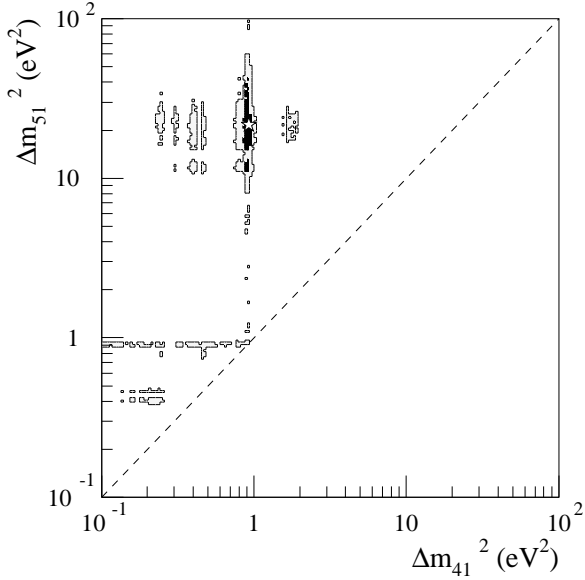}%
\caption{Allowed ranges in $(\Delta m_{41}^{2},\Delta m_{51}^{2})$ space for
(3+2) models, for the combined NSBL+LSND analysis, assuming statistical
compatibility between the NSBL and LSND datasets. The star indicates the
best-fit point, the dark and light grey-shaded regions indicate the 90 and
99\% CL allowed regions, respectively. }%
\label{3plus2scatter}%
\end{center}
\end{figure}

Fig.~\ref{3plus2scatter} shows the 90\% and 99\% CL allowed regions in
$(\Delta m_{41}^{2},\Delta m_{51}^{2})$ space obtained in the combined (3+2)
analysis. In light of the (3+1) analysis, the result is not surprising,
pointing to favored masses in the range $\Delta m_{41}^{2}\simeq
0.9\ \hbox{eV}^{2}$, $\Delta m_{51}^{2}\simeq10-40\ \hbox{eV}^{2}$, at 90\%
CL. At 99\% CL, the allowed region extends considerably, and many other
$(\Delta m_{41}^{2},\Delta m_{51}^{2})$ combinations appear. The best-fit
model ($\chi_{\hbox{\tiny SBL}}^{2}$=135.9, 145 d.o.f.) is described by the
following set of parameters: $\Delta m_{41}^{2}=0.92\ \hbox{eV}^{2}$,
$U_{e4}=0.121$, $U_{\mu4}=0.204$, $\Delta m_{51}^{2}=22\ \hbox{eV}^{2}$,
$U_{e5}=0.036$, and $U_{\mu5}=0.224.$ Solution with sub-eV neutrino masses are
also possible and the best fit value for these is given by: $\Delta m_{41}%
^{2}=0.46\ \hbox{eV}^{2}$, $U_{e4}=0.090$, $U_{\mu4}=0.226$, $\Delta
m_{51}^{2}=0.89\ \hbox{eV}^{2}$, $U_{e5}=0.125$, $U_{\mu4}=0.160$,
corresponding to $\chi_{\hbox{\tiny SBL}}^{2}$=141.5 (145 d.o.f.).%

\section{MiniBooNE and Sterile Neutrinos}

MiniBooNE will be one of the first experiments to check such sterile neutrino
models. The experiment will make a definitive investigation of the LSND
anomaly to determine if it is from neutrino oscillations and, if so, measure
the oscillation parameters. In addition, MiniBooNE will look for sterile
neutrinos through searches for a $\nu_{\mu}$ disappearance signal.

\subsection{The MiniBooNE Experiment}

\begin{figure}
[ptb]
\begin{center}
\includegraphics[
height=0.8363in,
width=3.039in
]%
{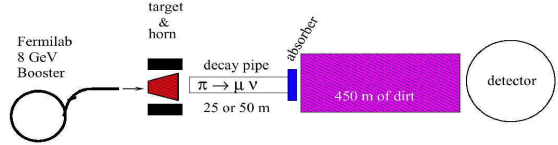}%
\caption{Schematic diagram of the MiniBooNE beamline. Protons from the
Fermilab 8 GeV Booster are focused onto a 1.7 interaction length beryllium
target. A magnetic horn then focuses positive (or negative) pions into a 50 m
long decay pipe. Following a steel dump at the end of the decay pipe is 490 m
of dirt shield before the MiniBooNE detector.}%
\label{MBBeam}%
\end{center}
\end{figure}

In order to test the LSND result, MiniBooNE is designed to sample an L/E
region similar to the LSND value of 1m/MeV while substantially changing the
sources and size of systematic errors. In addition, MiniBooNE will have
superior statistics allowing the experiment to make a definitive statement as
to whether oscillations exist in this region. To accomplish these goals, the
450-ton MiniBooNE detector is located at L = 541 m from the production target,
and the neutrino flux is produced from 8 GeV Fermilab Booster protons
impinging on a beryllium target giving neutrinos with a mean energy around 800
MeV. Compared to LSND, MiniBooNE neutrinos are produced from energetic (not
stopping) pions and the signature and backgrounds for $\overset{(-)}{\nu}_{e}$
appearance are much different. As shown in Figure \ref{MBBeam}, the neutrino
beam is produced by pion and kaon decays in a 50 m decay region downstream of
single, magnetic-horn focusing system. The proton beam arrives at the
production target within the horn during 1.6 $\mu$s beam spills that happens
at about 5 Hz.%

\begin{figure}
[ptb]
\begin{center}
\includegraphics[
width=3.039in
]%
{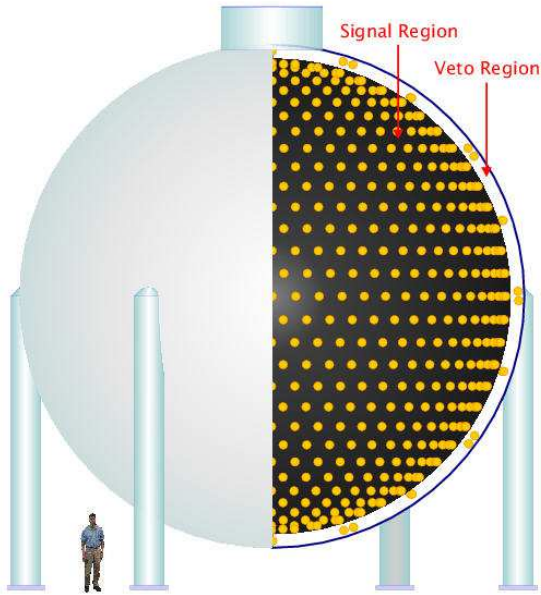}%
\caption{A schematic diagram of the MiniBooNE detector showing the signal and
veto regions. Each filled circle corresponds to a photomultiplier tube. The
veto region is isolated from the signal region by an optical barrier.}%
\label{MBDetector}%
\end{center}
\end{figure}

The MiniBooNE detector shown in Figure \ref{MBDetector} consists of a 12.2 m
diameter tank filled with pure mineral oil with the outer surface lined with
1280 photomultiplier tubes (PMTs). The fiducial volume corresponds to 450 tons
of oil and is defined by a veto region instrumented with 240 PMTs at the outer
radii. Neutrinos are identified and measured using the Cerenkov and the small
amount of scintillation light produced by outgoing charged tracks. The
interaction point, time of an event, and the directions of relativistic tracks
are reconstructed from the time and charge recorded in the PMTs.%

Particle identification is accomplished with a neural net that uses the
difference in characteristics of the Cerenkov rings and scintillation light
associated with electrons, muons, protons, or $\pi^{0}$'s. Most muon neutrino
events can be easily identified by their penetration into the veto region or
by their stopping and producing a Michel electron after a few microseconds. An
important background to the $\nu_{e}$ appearance search is $\nu_{\mu}$ neutral
current production of $\pi^{0}$'s, most of which are identified by the
production of two Cerenkov rings from the two decay gamma-rays.

Simple cuts requiring that a candidate neutrino event have a time within the
1.6$\mu$s beam spill, fewer than 6 veto hits, and more than 200 signal region
hits reduces cosmic ray background to 1 in 1000 with respect to beam events.
The beam that arrives at the detector is almost all $\nu_{\mu}$'s with a small
0.6\% contamination of $\nu_{e}$'s that mainly come from muon and kaon decay
in the decay pipe. Even though the kaon contribution is smaller than the muon,
the uncertainty in kaon production is a dominant systematic uncertainty in the
$\nu_{e}$ appearance oscillation measurement.

\begin{figure}
[ptb]
\begin{center}
\includegraphics[
height=2.0704in,
width=3.039in
]%
{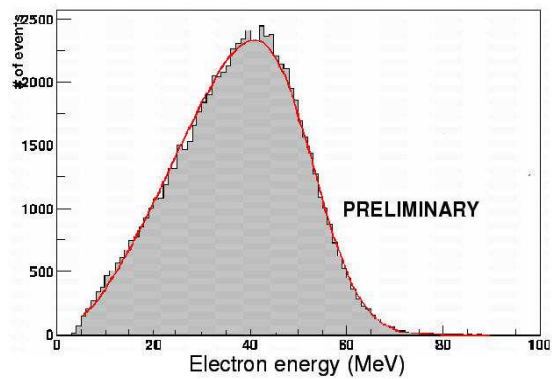}%
\caption{The Michel electron energy spectrum. The solid line is the expected
spectrum convolved with a Gaussian resolution factor. The distribution at the
endpoint indicates a 14.8\% energy resolution. }%
\label{michel}%
\end{center}
\end{figure}

Several methods are used to calibrate the detector over the full energy range
from 50 to 1000 MeV. The spectrum of observed Michel electrons from stopped
muon decay yields a calibration point at the 53 MeV Michel endpoint.
Figure~\ref{michel} shows good agreement between the data and expectation with
an energy resolution of 15\% at 53 MeV. Cosmic ray muons also provide an
energy calibration when their path length can be identified. To accomplish
this, a multi-plane, scintillator hodoscope on the top of the detector is
combined with scintillating cubes placed at various locations in the tank.
Using this combination, a series of muon ranges corresponding to energies of
100 to 1000 MeV can be isolated and used for calibration.%

\begin{figure}
[ptb]
\begin{center}
\includegraphics[
height=2.4958in,
width=3.039in
]%
{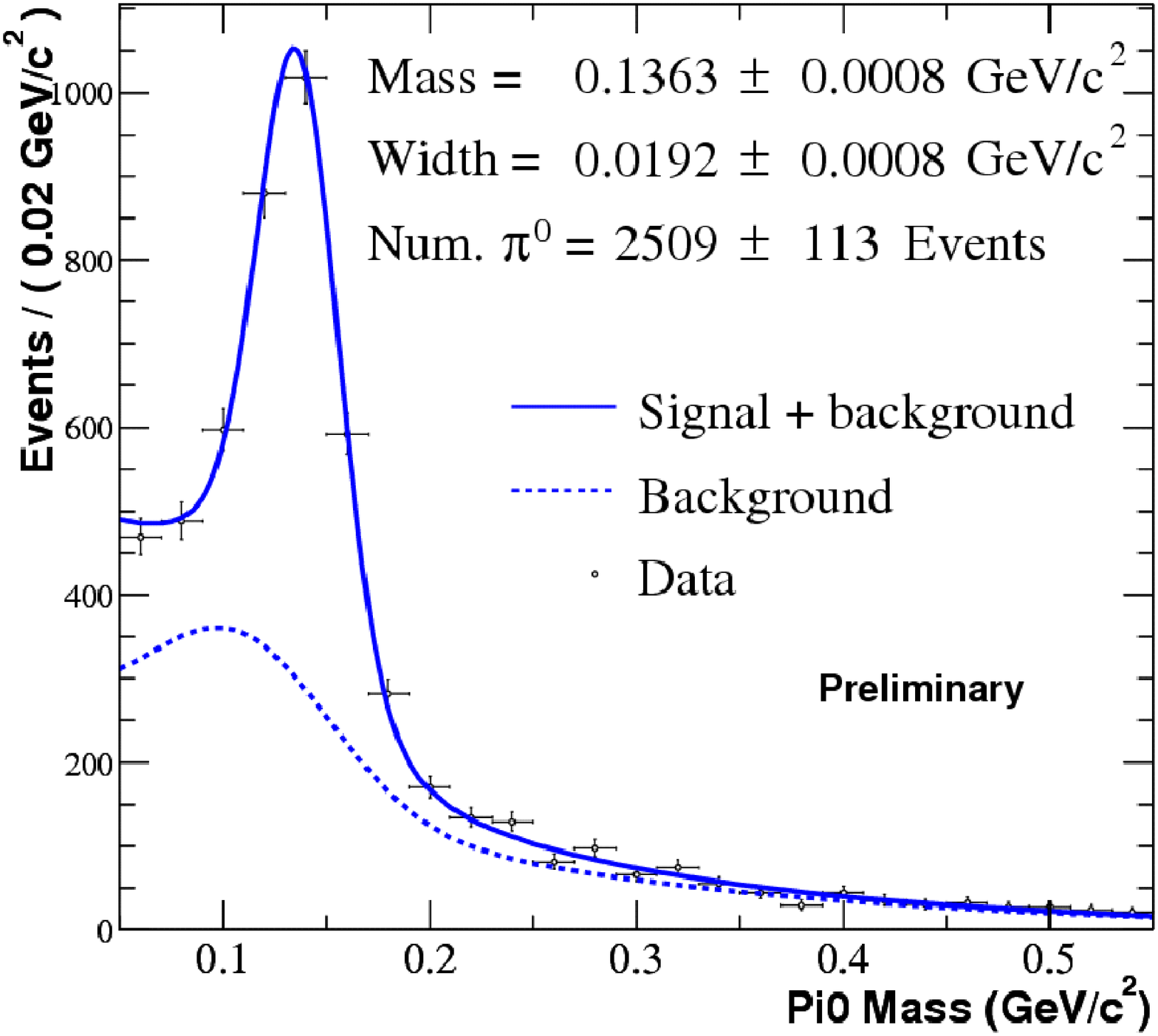}%
\caption{Reconstructed invariant mass for beam triggers satisfying "good
event" selection cuts plus a $>$ 40 MeV cut for each of the two assumed
Cerenkov rings. The data are shown by the points and the Monte Carlo
prediction for neutral current $\pi^{0}$ production and other backgrounds by
the solid and dashed curves, respectively. }%
\label{pi0mass}%
\end{center}
\end{figure}

Reconstructed $\pi^{0}$events provid yet another calibration source.
Figure~\ref{pi0mass} shows a preliminary invariant mass distribution for
events with two Cerenkov rings above 40 MeV. The photons that are included in
the pion mass distribution span a considerable range from 40 to over 1000 MeV.
The agreement of the mass peak at 136.3 $\pm$ 0.8 MeV with the 135.0 MeV
expectation then gives a check on the energy scale over this range of photon energies.%

\subsection{Oscillation Analysis: Status and Plans}

The MiniBooNE collaboration is committed to performing a \textquotedblleft
closed box\textquotedblright\ analysis for the $\nu_{e}$ appearance search.
For this procedure, the data events that could contain electron neutrinos in
the oscillation signal region are separated off and not available for full
detailed analysis. In this way, inadvertent biases that might be introduced
during the reconstruction development, tuning, and calibration can be avoided.
Data events in other categories such as muon and NC $\pi^{0}$ events are
available for study. The plan is to complete the full development,
calibration, and verification of the event reconstruction with Monte Carlo and
the \textquotedblleft open data\textquotedblright\ before considering the
\textquotedblleft closed box"\textquotedblright\ events.

The open data samples are already yielding interesting physics studies as well
as serving as a cross check of the data reconstruction. These studies include
a $\nu_{\mu}$ disappearance oscillation search, measurements of cross sections
for low-energy processes, studies of $\nu_{\mu}$ NC $\pi^{0}$ production, and
studies of $\nu_{\mu}$ NC elastic scattering.

A more detailed report on the status and plans for the MiniBooNE physics
analyses can be found in the \textquotedblleft MiniBooNE Run
Plan\textquotedblright\cite{MBRunPlan} that was submitted to Fermilab in
November 2003.

\subsubsection{$\nu_{\mu}$ Disappearance Search}

\begin{figure}
[ptb]
\begin{center}
\includegraphics[
height=2.9084in,
width=3.039in
]%
{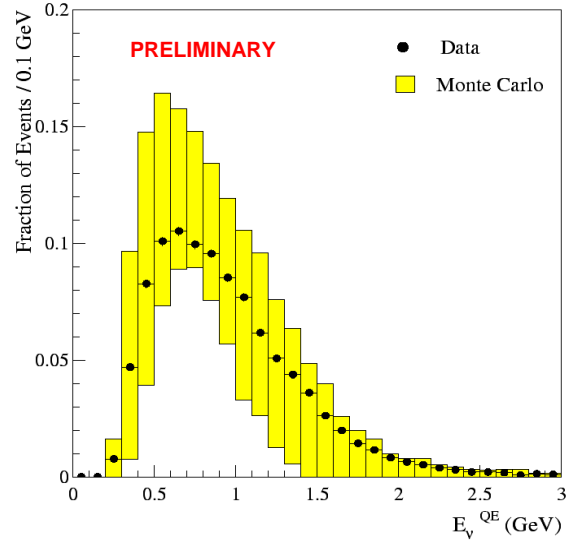}%
\caption{Reconstructed energy distribution for $\nu_{\mu}$ charged-current,
quasi-elastic events, data (black dots) and an example of the Monte Carlo
prediction with optical model variations. (boxes). The distributions are
nomalized to unit area and, thus, only display shape information.}%
\label{numuqe}%
\end{center}
\end{figure}

\begin{figure}
[ptb]
\begin{center}
\includegraphics[
height=2.2796in,
width=3.039in
]%
{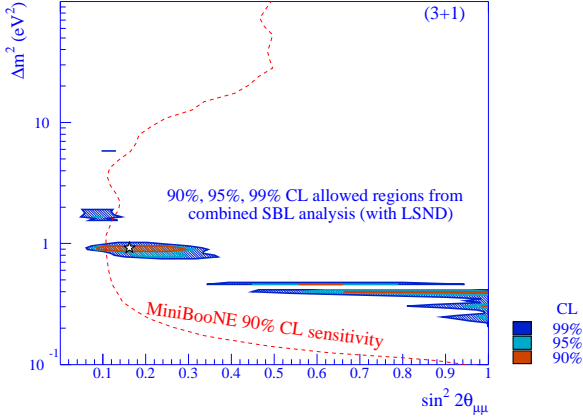}%
\caption{90\% CL sensitivity for a MiniBooNE $\nu_{\mu}$ disappearance search
with $1\times10^{21}$ protons on target. Also shown are the 90\%, 95\%, and
99\% CL regions from the (3+1) global fits to the NSBL and LSND data.}%
\label{disappear}%
\end{center}
\end{figure}

The $\nu_{\mu}$ disappearance search mainly uses quasi-elastic,
charged-current (CC) events, $\nu_{\mu}+n\rightarrow\mu^{-}+p$. These events
are isolated by identifying a single ring topology and PMT hit timing cuts.
For quasi-elastic events, the incoming neutrino energy can be simply estimated
from the energy and angle of the outgoing muon (or electron) using the
formula:%
\[
E_{\nu}^{QE}=\frac{1}{2}\frac{2ME_{\mu}-m_{\mu}^{2}}{M-E_{\mu}+p_{\mu}%
\cos\theta_{\mu}}%
\]
From Monte Carlo studies, this procedure yields an energy resolution of about
10\% at 800 MeV. With only a single detector, the oscillation search is
performed by comparing the shape of the energy distribution to the Monte Carlo
expectation. The Monte Carlo expectation has inputs from measured pion
production experiments for energies around 8 GeV and from quasi-elastic cross
section measurements. With an early data sample of 30,000 $\nu_{\mu}$ CC
quasi-elastic events, Figure~\ref{numuqe} shows a comparison of the measured
quasi-elastic energy, $E_{\nu}^{QE}$, distribution in the data to Monte Carlo
(each normalized to unit area). The error bars on the Monte Carlo include the
current assessment of the major sources of systematic uncertainties. In
particular, the errors include uncertainties associated with the $\nu_{\mu}$
flux, $\nu_{\mu}$ CC quasi-elastic cross section, and the properties of light
production and transmission in the MiniBooNE detector. Substantial reductions
of these uncertainties are expected as the analysis progresses. With the full
data sample of several hundred thousand $\nu_{\mu}$ CC quasi-elastic events,
the estimated sensitivity to a disappearance signal is shown in
Figure~\ref{disappear} along with the allowed regions from the previously
described (3+1) sterile neutrino model fits. As that figure shows, the
MiniBooNE measurement will address important regions for the (3+1) and
therefore also the (3+2) models.%

\subsubsection{The $\nu_{\mu}\rightarrow\nu_{e}$ Appearance Search}

\begin{figure}
[ptb]
\begin{center}
\includegraphics[
width=3.039in
]%
{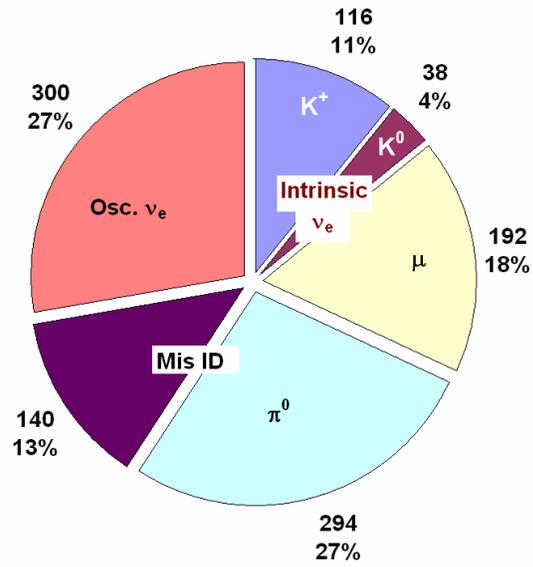}%
\caption{Statistics for various signal and background event categories. The
numbers are for $1\times10^{21}$ protons on target and are based on the
observed neutrino flux. The number of oscillation events as shown is for the
mid-range of the LSND allowed region and may vary by about 20\% with the
specific $\Delta m^{2}$ value. }%
\label{events}%
\end{center}
\end{figure}

A prime goal of MiniBooNE is to make a definitive search for neutrino
oscillations in the LSND parameter region by looking for an anomalous signal
of $\nu_{e}$ appearance over background. The beam and detector have been
designed to cover this region with good $\nu_{e}$ efficiency while minimizing
backgrounds. An important feature of the MiniBooNE design is to constrain all
backgrounds through either internal or external data measurements as outlined
below. For $1\times10^{21}$ protons on target, the expected number of
oscillation and background events is given in Figure~\ref{events}.%

One of the major sources of background events is the intrinsic electron
neutrinos in the beam. This background $\nu_{e}$ flux comes from muon and kaon
decay and is at the $6\times10^{-3}$ level with respect to the $\nu_{\mu}$
event rate. For the $\nu_{e}$'s from $\mu$-decay, the flux is directly tied to
the observed $\nu_{\mu}$ flux, since the MiniBooNE detector only subtends
neutrinos from very forward pion decays. The pion energy distribution can then
be determined from the observed $\nu_{\mu}$ events and used to predict the
$\nu_{e}$'s from $\mu$-decay. The other major source of $\nu_{e}$'s is from
$K_{e3}$ decay. This source will be determined using data from two low energy
production experiments that have recently taken data, the HARP experiment at
CERN\cite{HARP} and the E910 experiment at BNL\cite{E910}. With these
production data, it is estimated that the $\nu_{e}$ flux from kaon decay can
be determined with about a 5\% uncertainty. An additional check will also be
available from a special system, the Little Muon Counters, that measures the
decay muons from pion and kaon decay coming at wide angles from the decay pipe.

\begin{figure}
[ptb]
\begin{center}
\includegraphics[
height=2.0755in,
width=3.0381in
]%
{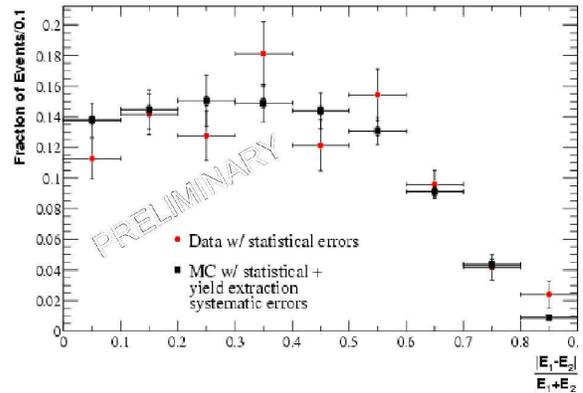}%
\caption{Fraction of NC single $\pi^{0}$ events as a function of the asymmetry
of the reconstructed gamma energies. The event fraction in each bin is
extracted via a fit to the invariant mass plot for events in that bin. The
background and \textquotedblleft signal\textquotedblright\ shapes used in the
fit are from a MC-based parameterization.}%
\label{pi0asym}%
\end{center}
\end{figure}

The other major source of background events is NC $\pi^{0}$ production, where
one of the gammas from the $\pi^{0}$ decay overlaps the other or is too low
energy to be detected. Over 99\% of the NC $\pi^{0}$ events are rejected in
the appearance analysis, and about 22\% are fully reconstructed, which allows
the overall number and kinematic distributions to be studied. As seen in
Figure~\ref{pi0mass}, the invariant mass distribution is understood very well.
The identified $\pi^{0}$ events in the data can then be used to extrapolate
into the kinematic region of the background. As an example,
Figure~\ref{pi0asym} shows the asymmetry distribution of the two photons for
identified $\pi^{0}$ events for data and Monte Carlo; the good agreement for
events with large asymmetry gives confidence that this background can be understood.%

\begin{figure}
[ptb]
\begin{center}
\includegraphics[
height=3.109in,
width=3.039in
]%
{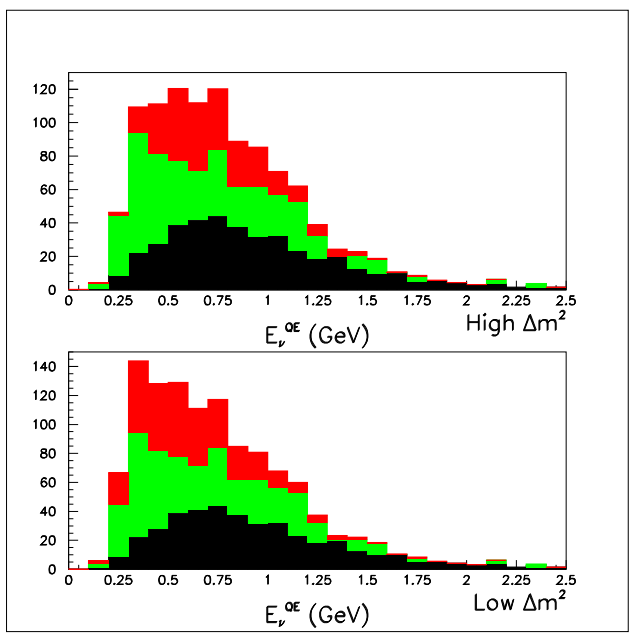}%
\caption{Summed event energy distribution for the oscillation signal and
backgrounds. The bottom black area is the intrinsic $\nu_{e}$ backgound
events, the middle light region is the NC $\pi^{0}$ events, and the top area
is the oscillation events for 1.0 (top) and 0.4 (bottom) eV$^{2}$. The data
sample corresponds to $1\times10^{21}$ protons on target.}%
\label{evtEdist}%
\end{center}
\end{figure}

\begin{figure}
[ptb]
\begin{center}
\includegraphics[
height=3.8043in,
width=3.039in
]%
{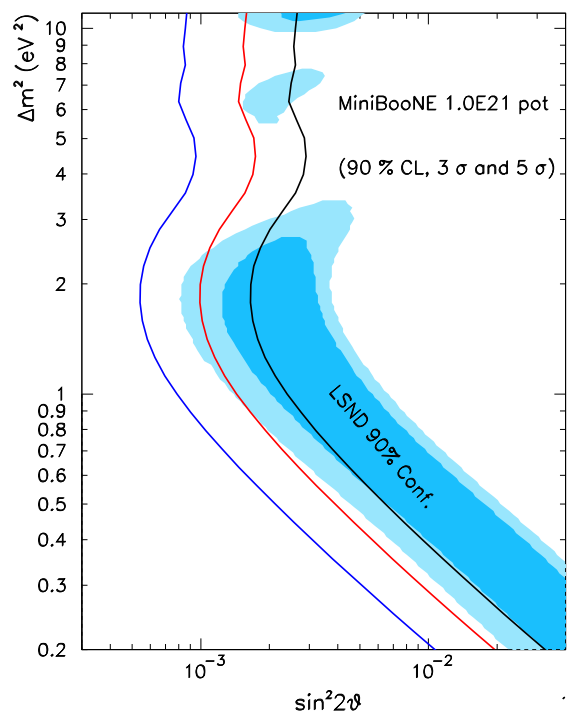}%
\caption{Estimate of the MiniBooNE oscillation sensitivity for $1\times
10^{21}$ protons on target using fits to the event energy distribution
including signal and backgrounds. The dark (light) areas are the LSND 90\%
(99\%) CL allowed regions. The three curves give the 90\%, 3 $\sigma$, and 5
$\sigma$ sensitivity regions for MiniBooNE. }%
\label{osclimit}%
\end{center}
\end{figure}

\begin{figure}
[ptb]
\begin{center}
\includegraphics[
height=3.8138in,
width=3.039in
]%
{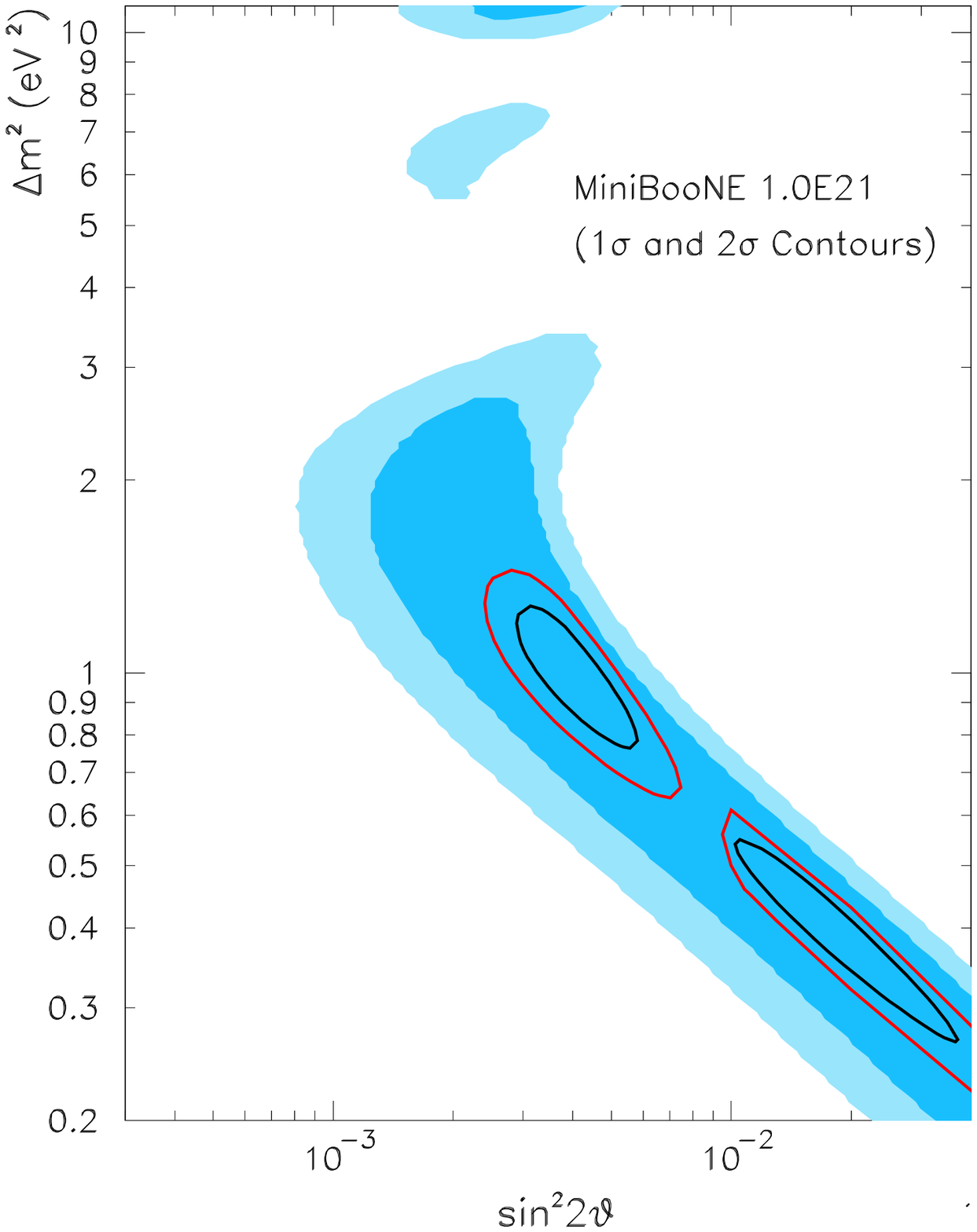}%
\caption{One and two sigma contours for an oscillation signal with $\Delta
m^{2}$ = 0.4 or 1.0 eV$^{2}$ for a data sample corresponding to $1\times
10^{21}$ protons on target. }%
\label{oscsignal}%
\end{center}
\end{figure}

Another handle on the background is the energy distribution of the various
components as shown in Figure~\ref{evtEdist} where the background is summed
with a signal for a high (low) $\Delta m^{2}$ value of 1 (0.4) eV$^{2}$. The
intrinsic $\nu_{e}$ events show a clear high energy tail which can be
identified and the misidentified $\pi^{0}$ events are narrower in energy than
an oscillation signal, so a fit to the energy distribution can be used to
enhance the sensitivity to an oscillation signal. The MiniBooNE sensitivity
has been estimated by performing such energy-dependent fits to simulated data
with no underlying oscillation signal. Figure~\ref{osclimit} shows the
expected 90\%, 3$\sigma$, and 5$\sigma$ MiniBooNE exclusion regions for an
underlying null signal; Figure~\ref{oscsignal} shows how well an oscillation
signal can be measured at either low or high $\Delta m^{2}$.%

At the current time MiniBooNE has collected about $2.5\times10^{20}$ protons
on target. The data collection rate is steadily improving as the Booster
accelerator losses are reduced. Many improvements are being implemented into
the Booster and LINAC which not only help MiniBooNE but also the Tevatron and
NuMI experiments. The MiniBooNE collaboration plans to \textquotedblleft open
the $\nu_{e}$ appearance box\textquotedblright\ when the analysis has been
substantiated and when sufficient data has been collected for a definitive
result; this is expected to be sometime in 2005.

\subsection{Future Plans and Possible Follow-up Experiments}

\begin{figure}
[ptb]
\begin{center}
\includegraphics[
height=3.1903in,
width=3.039in
]%
{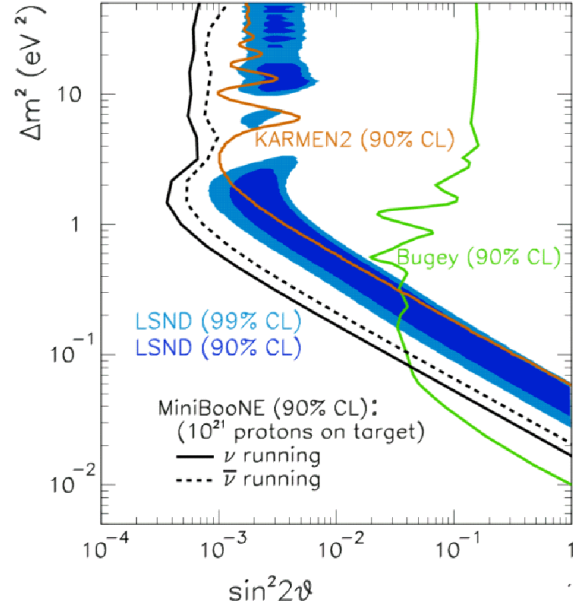}%
\caption{Estimate of the MiniBooNE oscillation sensitivity at 90\% CL for
$1\times10^{21}$ protons on target for neutrino and antineutrino running. The
dark (light) areas are the LSND 90\% (99\%) CL allowed regions. Also, shown
are the 90\% CL limits from the KARMEN2 and Bugey experiments.}%
\label{osclimit_nubar}%
\end{center}
\end{figure}

Whether or not MiniBooNE sees an oscillation signal in neutrino running mode,
it is also important to run with antineutrinos. Antineutrino results will both
directly address the LSND anomaly, which was associated with a $\overline{\nu
}_{\mu}\rightarrow\overline{\nu}_{e}$ signal; and allow investigations of
possible CP or CPT violation effects. If MiniBooNE sees a signal in neutrino
mode, then it will be important to see if the mixings and $\Delta m^{2}$
values are the same for incident antineutrinos. The antineutrino event rate is
about a factor of four lower than the neutrino rate for a given number of
protons on target, but the $\pi^{0}$ and intrinsic $\nu_{e}$ backgrounds are
significantly lower. Because of these effects, one needs to run twice as long
in antineutrinos to reach similar sensitivity as for neutrino running as shown
in Figure~\ref{osclimit_nubar}.%

\begin{figure}
[ptb]
\begin{center}
\includegraphics[
height=2.028in,
width=3.039in
]%
{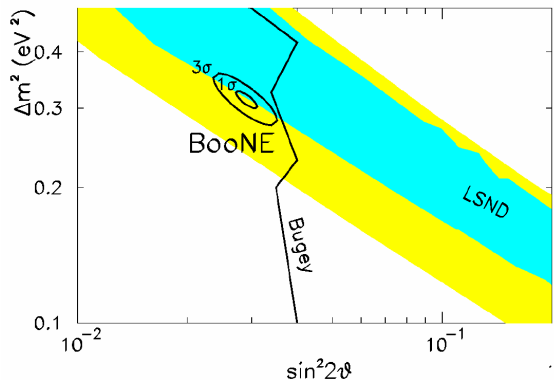}%
\caption{Example of a two detector BooNE measurement for an underlying
oscillation scenario with $\Delta m^{2}=0.3$ eV$^{2}$. The one and three sigma
regions are displayed.}%
\label{boonesignal}%
\end{center}
\end{figure}

If MiniBooNE sees an oscillation signal, the planned next step is to add a
second detector at an optimal location (2 km for low $\Delta m^{2}$ or 0.25 km
for high $\Delta m^{2}$) in order to explore the oscillation parameters; this
upgrade is referred to as the \textquotedblleft BooNE\textquotedblright%
\ experiment. This two (or more) detector setup would allow precision
measurements of the parameters, $\Delta m^{2}$ to $\pm0.014$ eV$^{2}$ and
$\sin^{2}2\theta$ to $\pm0.002$. Figure~\ref{boonesignal} shows an example
measurement for an underlying $\Delta m^{2}$ value of 0.3 eV$^{2}$. With this
sensitivity, there are many questions concerning the oscillation signal that
the BooNE experiment will explore. The first question to address is if the
signal is consistent with the L/E behavior expected for neutrino oscillations.
In the (3+n) type sterile neutrino models, BooNE can determine how many
separate $\Delta m^{2}$ there are. Finally, with two detectors, BooNE can
perform a much more precise $\nu_{\mu}$ disappearance search that would
directly probe for indications of oscillations to sterile neutrinos.%

A positive MiniBooNE signal will also prompt investigations by other
experiments at Fermilab, BNL, CERN, and JPARC. It will be important to map out
the oscillations related to all three types of active neutrinos, $\nu_{e},$
$\nu_{\mu},$ and $\nu_{\tau}$, in this high $\Delta m^{2}$ region with both
appearance and disappearance measurements. A $\nu_{\mu}\rightarrow\nu_{\tau}$
appearance experiment using, for example, an Opera-style emulsion detector
could explore this region but would require a higher energy neutrino beam to
overcome the $\tau$ production threshold effects; the NuMI 8 GeV medium energy
beam is a possibility.

In conclusion, neutrinos have provided many surprises over the past decade and
will most likely continue to do so. Although the \textquotedblleft neutrino
standard model\textquotedblright\ can be used as a guide, the future direction
for the field will be determined by what we discover from experiments. The
possibility that sterile neutrinos exist may open up a whole new area to explore.

\end{document}